\documentclass[12pt,aps]{revtex4}
\usepackage{epsfig,amsmath,amssymb}

\newcommand{\bea}{\begin{eqnarray}}
\newcommand{\eea}{\end{eqnarray}}

\begin{document}

\title{Colored  scalars   and  the neutron electric dipole moment}

\author{Svjetlana Fajfer}
 \email[Electronic
address:]{svjetlana.fajfer@ijs.si} 
\affiliation{Department of Physics, University of Ljubljana, Jadranska 19, 1000 Ljubljana, Slovenia}
\affiliation{J. Stefan Institute, Jamova 39, P. O. Box 3000, 1001
  Ljubljana, Slovenia}

\author{Jan O. Eeg} 
\email[Electronic address:]{j.o.eeg@fys.uio.no} 
\affiliation{Department of Physics, University of Oslo, P.O.Box 1048 Blindern, N-0316 Oslo, Norway}

\date{\today}

\begin{abstract}
We investigate new contributions to the neutron electric  dipole moment
 induced by colored scalars.
 As an example,  we evaluate  contributions  coming from  the color octet, 
weak doublet  scalar, accommodated within  a modified Minimal Flavor Violating 
framework. These flavor non-diagonal couplings of the color octet scalar  
might   account for the measured assymmetry 
$a_{CP} (D^0 \to K^- K^+) -  a_{CP} (D^0 \to  \pi^+ \pi^-)$
at tree level.
 The same  
couplings constrained by this assymmetry also induce 
 two-loop contributions  to the neutron electric  dipole moment.  
We find that the direct CP violating asymmetry in neutral  $D$-meson decays 
is more constraining on the allowed parameter space than the  current experimental bound  on neutron electric dipole moment.
We comment also on contributions 
of higher dimensional operators to the neutron electric dipole moment 
 within this framework.

\end{abstract}

\maketitle

\section{Introduction}
The neutron electric dipole moment (NEDM) plays a special role in  current 
searches of  physics  beyond the Standard Model (SM). The low energy
 flavor physics puts extremely tight bounds on possible non-standard model
 contributions to the NEDM, and in particular special attention has been paid
 to  new sources of CP violation.  The NEDM gives a great
 opportunity to learn about additional sources of CP violation. 
 There are many studies on the NEDM
 (for a  review see \cite{Pospelov:2005pr}).
 In addition to CP violating effects, the
  NEDM   still offers many puzzles for the study of non-perturbative
 QCD effects within the SM. 

 Measurements of the CP asymmetry  in $D \to K^+ K^-/\pi^+ \pi^-$
 larger than SM expectations have attracted many theoretical studies. 
The LHCb collaboration recently updated  their analysis  leading to a 
  decreased value of the world average CP asymmetry
 \cite{HFAG,Aaij:2011in,Aaltonen:2011se}:
\begin{equation}
\Delta a_{CP} = (-0.329 \pm 0.121) \%\,, 
\label{e1}
\end{equation}
with $\Delta a_{CP} = a_{K^+K^-} - a_{\pi^+ \pi^-}$ and the definition 
\begin{equation}
a_f \equiv \frac{\Gamma(D^0 \to f) - 
\Gamma(\bar D^0 \to f) }{\Gamma(D^0 \to f) + \Gamma(\bar D^0 \to f) }\,.
\label{e2}
\end{equation}
Many theoretical studies were performed in order to explain 
the apparent discrepancy 
\cite{Isidori:2011qw,Grossman:2006jg,Cheng:2012wr,Li:2012cfa,
Franco:2012ck,Pirtskhalava:2011va,Bhattacharya:2012ah,Brod:2012ud,
Feldmann:2012js,Cheng:2012xb,Lenz:2013pwa}. 
 Some of these approaches
  have  explained the 
 observed asymmetry by 
SM  effects \cite{Brod:2012ud,Cheng:2012xb}, as already indicated some time ago
\cite{Golden:1989qx}, while others have considered, 
possible new physics (NP).   In ref. \cite{Brod:2012ud} the authors
 argued that penguin contraction power corrections
might significant enhance the decay amplitudes. They also found
  that the same mechanism might consistently explain  the branching ratios
 for  singly Cabibbo-suppressed $D \to PP$ decays. 
 The authors of  ref. \cite{Isidori:2011qw}
pointed out that most likely the
effective operators explaining CP asymmetry in charm decays are 
  color-magnetic dipole operators, although  color octet scalar as well 
as two Higgs doublet models in specific parameter space are still allowed
 options \cite{Altmannshofer:2010ad}.
 The important result of these studies
 \cite{Isidori:2011qw,Grossman:2006jg,Li:2012cfa,Lenz:2013pwa} 
 is that apparently one needs an additional source 
of CP violation and in particular in the charm sector. 
On the other hand, CP violation in the $B$ system has been related to
the  NEDM \cite{Buras:2010zm}, while in \cite{Altmannshofer:2010ad} 
CP violation  in $D^0~-~\bar D^0$ oscillation was related to the NEDM. 
In the study of the NEDM usually the lowest dimensional operators were 
considered \cite{Pospelov:2005pr}. Recently, the authors
 of \cite{Mannel:2012qk} found that  higher dimensional operators 
might lead to rather important contributions to NEDM, as noticed some 
time ago in \cite{Eeg:1982qm,Eeg:1983mt}.
  The authors of   \cite{Mannel:2012qk} also
noticed   \cite{Mannel:2012hb} that CP violation in charm decays  
 can induce an  increase of the NEDM, if these higher dimensional operators
 are included. It should also be noted that several authors have
 recently considered the NEDM within frameworks beyond the SM;- see for 
instance refs. \cite{Brod:2013cka,He:2014uya,Gorbahn:2014sha}. 

Experimental results agree remarkably well with SM predictions in the case 
of the $K$ and $B_{d,s}$ systems.  
The Minimal Flavor Violation (MFV) 
framework \cite{D'Ambrosio:2002ex,Buras:2010zm}, in which  flavor-changing
 transitions in the quark sector
 are entirely described 
by  two quark Yukawa couplings, has been used to parametrize NP effects. 
The MFV principle does not  forbid appearance  of new flavor blind
 CP violating phases in addition to the unique phase of 
the CKM matrix. The Flavor Changing Neutral Currents (FCNC) coming 
from down-like quark sectors are very well known and constrained
 by experiment. In addition to the well understood 
$s \to d$ transition within $K$ meson physics, rare $B$ decays are 
excellent probes of NP beyond the SM.
 The recent model independent study of ref. \cite{Altmannshofer:2012az} 
has been aimed to constrain NP
 contributions in rare $B$ meson decays.  Among the relevant observables 
used in $B_s \to \mu^+ \mu^-$, $B\to K \mu^+ \mu^-$, 
 $B\to K^* \mu^+ \mu^-$, and  $B \to X_s \gamma$  decays, the measured 
CP asymmetry in  $B \to X_s \gamma$ is specially informative.
Although the hadronic uncertainties are still large,  
 the $A_{CP} (B \to X_s \gamma)$ might  constrain 
CP violating NP contributions.  Namely, the measured branching ratios  
are not itself constraining enough for the possible CP violating NP effects.
 The strongest bound  on new CP violating parameters  comes from
$ b\to s \gamma$ decay  \cite{HFAG-b}:
$ A_{CP}^{exp}(b\to s \gamma) = (-1.2\pm 2.8)\%$.

Using the constraint from charm decays (\ref{e1}), 
 we consider contributions of the additional heavy colored scalar meson 
 to the NEDM.
 As a particular example we consider a color octet, weak
 doublet scalar introduced in ref. \cite{Manohar:2006ga}. These authors
 have  analyzed a model  with the most general scalar structure
of the standard model (SM). 
  The main request on the  model in \cite{Manohar:2006ga} was that  
it maintains the smallness of  flavor changing neutral currents (FCNC),
 or in another words it is a model which fulfills principles of MFV.
 The new scalar state might 
contain a new source of CP violation. The  phenomenology of SM
 modified by the presence of the color octet  scalar state was investigated
 in \cite{Manohar:2006ga,Gresham:2007ri}. 
 The constraints on  of the color octet scalar coupling to up-type of
 quarks that arises from precision electroweak data on 
$R_b$ in $Z\to b \bar b$  decays, have been considered
 in \cite{Gresham:2007ri}.  This study is not constraining the mass 
of the color-octet.
 The NEDM gets its largest additional contribution within the 
proposed model \cite{Manohar:2006ga}
by the color electric dipole moment (CEDM) of the $b$-quark.  When
 integrated out, the  CEDM of
the  $b$-quark \cite{Boyd:1990bx} induces  
Weinberg's three-gluon CP violating operator \cite{Weinberg:1989dx}. 
 The electric dipole operator of the $d$-quark  also induces a NEDM,  
as discussed in \cite{Arnold:2012sd}.
Namely, the existence of two different
 couplings of   $d$-quark with the u-quark types  lead to a 
one-loop penguin-like contribution, 
and it can then be present in the
 case of $d$-quark EDM.

The effective Lagrangian for a dipole moment of a fermion $f$ has 
the generic form
\begin{equation}
{\cal L}_\mathrm{eff} \; = \; \frac{i}{2} \, d_f \,
\bar{f} \sigma_{\mu \nu} \, F^{\mu \nu} \, \, \gamma_5 \, f \; \, ,
\label{effLag}
\end{equation}
where $d_f$ is the EDM
 of the fermion,
$f$ is the fermion field, $F^{\mu \nu}$ is the electromagnetic field and
$\sigma_{\mu \nu} = i [\gamma_\mu, \, \gamma_\nu]/2$. 
 Contributions to  the NEDM come from 
 EDMs of single quarks, and contributions due to interplay of quarks 
within the neutron.
In the valence approximation, the contributions to NEDM from EDMs  
of single quarks ($d_q$)
are given by
\begin{equation}
d_n \, = \, \frac{4}{3} \, d_d \; - \; \frac{1}{3} \, d_u \; .
\label{valence}
\end{equation}
The existing experimental bound  from ref. \cite{Baker:2006ts} 
for the NEDM is:
\begin{equation}
d_n^{exp} \le 2.9\times 10^{-26} e \enspace {\rm  cm} \; ,
\label{e3}
\end{equation}
which bounds the appropriate imaginary parts of  couplings of quarks
 with colored scalars. 
Motivated by the  explanation of direct CP violation in 
decay of the neutral $D$-meson  by  the presence of a  
color octet,  we investigate the impact of the  imaginary couplings  
of quarks to the color scalars on the NEDM.
The color octet scalar can be searched at hadron colliders in
di-jet events \cite{Manohar:2006ga,Bertolini:2013vta,Cao:2013wqa,Trott:2010iz}.
 Current experimental searches at LHC based on the dijet analysis at 
ATLAS   \cite{ATLAS:2012ds}  and CMS \cite{CMS:2012yf}, as well as 
former at Tevatron excludes existence of a color octet scalar with
 mass below $1.86$ ${\rm TeV}$, while  four-jet  CMS searches
 do not observe it at a low-energy region \cite{KaiYifortheCMS:2013vqa} 
from $250-740$ ${\rm GeV}$. 
Also, ATLAS searches exclude it in 
low energy region  \cite{ATLAS:2012pu}. 
However, if the color octet scalar decays in more than two light quarks
(two jets), top-quark and jet, or $t \bar t$, then the existing 
bound on the color octet scalar mass would be different.  Then
 even masses of
order 400 GeV can not be excluded, as noticed in \cite{Heng:2013cya}. 
In Sec. II we first remind on
 the NEDM within the  SM. Sec. III is devoted to the study of
 NEDM contributions
obtained within a
  modified  color octet model \cite{Altmannshofer:2012ur} which 
  produces CP violation in charm at
 tree level. In Sec. IV we discuss obtained results.\\

 \section{NEDM in the SM}

\begin{figure}[t]
\begin{center}
    \epsfig{file=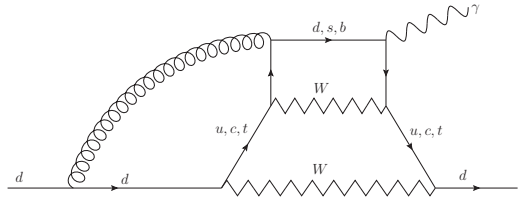,width=7cm}
\caption{Typical diagram for an EDM for  single quark within the SM.}
\label{fig:SM3loop}
\end{center}
\end{figure}

The NEDM has been studied for many years, both within and beyond the
 SM (for a review see \cite{Pospelov:2005pr}).
As it turned out that EDMs 
in the SM were small,
calculations within  new physics scenarios 
 were also performed.
To obtain a CP violating amplitude within the SM,
two weak interactions are needed, and   at least one of these
must be  a  penguin-like interaction.
 The EDMs of single quarks, which are three loop diagrams with double 
GIM-cancellations and at least one gluon exchange, 
are  of order  
 $\alpha_s \, G_F^2$ and proportional to quark masses and 
an imaginary CKM factor. A typical diagram is shown
 in Fig.~\ref{fig:SM3loop} .
However, in the SM, EDMs of single quarks were found to be very small, 
 namely of order $10^{-34} \, e$ cm according to studies of 
refs. \cite{Shabalin:1980tf,Czarnecki:1997bu}. 

\begin{figure}[t]
\begin{center}
    \epsfig{file=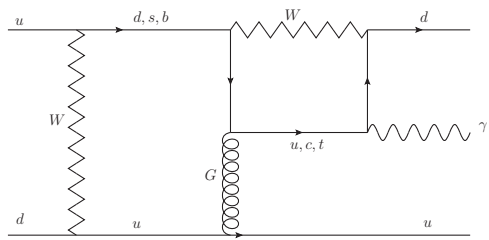,width=7cm}
\caption{Two-loop diagram within SM giving the diquark mechanism for NEDM.}
\label{fig:SM2loop}
\end{center}
\end{figure}

It was shown that a mechanism with interplay of weak and strong interactions
would give a nonzero contribution. Typically, such  amplitudes were written as
 baryon poles with two weak interactions, an ordinary W-exchange
 and a CP-violating penguin interaction, with for instance  a negative
 parity strange baryon as intermediate baryon, and a soft photon
 emitted from somewhere \cite{Nanopoulos:1979my,Morel:1979ep,Gavela:1981sk}.
There were also contributions  due to amplitudes at baryonic/mesonic  level
 where the  photon was emitted
 from an intermediate pion or kaon within a chiral loop
\cite{Khriplovich:1981ca,McKellar:1987tf}.
Within such mechanisms,  the $d_n/e$ was estimated to
be of order $10^{-33}$ to $10^{-31}$ cm.
It was pointed out that if the pole diagrams were interpreted at  quark level it would correspond  to   
a ``diquark mechanism'' \cite{Eeg:1982qm,Eeg:1983mt}. 
This means  a CP-violating 
two loop diagram for the quark process
 $u \, d \, \rightarrow \, d \, u \, \gamma$, 
as shown in Fig.~\ref{fig:SM2loop}.
Here the contribution was obtained in terms of a two loop factor
 proportional to $\alpha_s \, G_F^2$,and  a CKM factor, written as 
$F_{CKM} \; = \; Im\left(V_{ub} \, V_{tb} \, V_{td}, V_{ud} \right)$. 
 Such diagrams  lead to an eight dimensional operator for the NEDM  
  given by 
\begin{equation}
   Q_{di-quark} \, = \, F_{\alpha \beta} \, \epsilon^{\alpha \beta \mu \nu} \, 
\bar{u}_L \gamma_\mu  d_L \; \bar{d}_L \gamma_\nu  u_L \; ,
\label{edq}
\end{equation}
where $F_{\alpha \beta}$ is the electromagnetic tensor. 
The matrix element of this operator is found to be of order 
(1  -  6)$\times 10^{-3} \times m_N^3$, where $m_N$ is the nucleon mass.
The obtained result is 
$d_n/e \sim 10^{-33}$
to $10^{-32}$ cm \cite{Eeg:1982qm,Eeg:1983mt,Hamzaoui:1985ri}. 
Within this mechanism the NEDM
 is suppressed by a small
  hadronic matrix element, but had logarithmic GIM instead of power-like, 
compensating for hadronic suppression.

\begin{figure}[t]
\begin{center}
    \epsfig{file=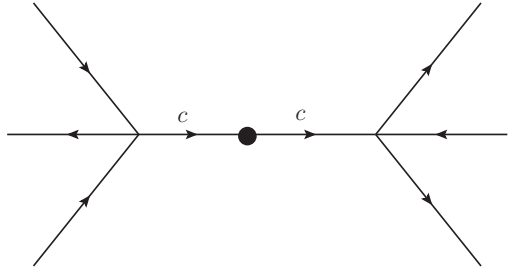,width=7cm}
\caption{Eight-dimensional contribution in the SM. The black dot on 
the $c$-quark propagator denotes the static contribution proportional
 to $1/m_c$.}
\label{fig:loopless}
\end{center}
\end{figure}

Using the  operator product expansion technique, one might consider
 operators of  higher dimensions. Recently, it was  
found \cite{Mannel:2012qk} that a tree-level higher dimensional
 operator might give a significant contribution to the NEDM comparable to the 
  loop induced  contributions. 
 Their contribution is illustrated in Fig. \ref{fig:loopless}. 

 The operator in (\ref{effLag}) has  dimension  five  
and it is the lowest dimension operator in SM, if the  $\theta$
 problem in QCD is rotated away as described in ref. \cite{Pospelov:2005pr}. 
If the electromagnetic field is replaced by gluonic field
 $G_{\mu \nu}^a$ then one has also CEDMs
 of the same dimension. 
However, operators of higher dimension, as the well-known 
 Weinberg operator \cite{Weinberg:1989dx,Jung:2013hka,Demir:2002gg}
\begin{equation}
{\cal L}_\mathrm{eff}^W \; = \;   \frac{C_W}{6} \, f^{abc} \, 
G_{\mu \nu}^a \epsilon^{\nu \beta \rho \sigma} \,G_{\rho\sigma}^ b\,G^{\mu c}_{\beta} 
\, ,
\label{effLag-W}
\end{equation}
contribute significantlly to the NEDM \cite{Pospelov:2005pr,Manohar:2006ga}.
An additional dimension six operator is 
\begin{equation}
{\cal L}_\mathrm{eff}^{ff^\prime}\; = \;  C_{ff^\prime} ( \bar \psi_f  \psi_{f}) \,
   (\bar \psi_{f^\prime}  i \gamma_5  \psi_{f^\prime}) \; ,
 \label{effLag-ff}
\end{equation}
which has been considered in ref. \cite{Pospelov:2005pr,Buras:2010zm}.
There are also so-called Barr-Zee contributions \cite{Barr:1990vd} on
 the two loop level contributing  to the NEDM
 within the SM, being 
competitive in  size  with  the Weinberg's three-gluon operator. 

 \section{Contributions from the color octet 
scalars within modified MFV framework}

The  extension of the SM introduced in \cite{Manohar:2006ga,Degrassi:2010ne} 
allows couplings of a color octet, weak doublet state with 
 weak  hyper-charge $1/2$ $(8,2)_{1/2}$ to quarks written as:
\begin{eqnarray} 
{\cal L}&=& - \sqrt 2 \eta_U \bar u_R^i \frac{m_U^i}{v} T^A u_L^i  \phi^{A0}
  +  \sqrt 2 \eta_U \bar u_R^i \frac{m_U^i}{v} T^A  V_{ij} d_L^i  \phi^{A+}
 \nonumber\\
&-& \sqrt 2 \eta_D \bar d_R^i   \frac{m_D^i}{v} T^A d_L^i  \phi^{A0\dag} \, - \,
 \sqrt 2 \eta_D \bar d_R^i \frac{m_D^i}{v}  T^A  V_{ij}^{\dag} u_L^i  \phi^{A-}
\,  +  \, h.\, c.\,.
\label{e4}
 \end{eqnarray} 
The couplings $\eta_{U,D}$ are universal complex numbers,
 $T^A$ are color $SU(3)$
 generators,  lower case roman
letters denote mass eigenstate fields and $\phi^{A+}$ and $\phi^{A0}$ are 
charged and neutral component of the scalar color octet. Further, $v$ is
 the vacuum expectation value (VEV) of  the SM Higgs doublet 
 and $m_{U,D}^i$ are quark masses. 
The  scalar
 octet state as a weak doublet might slightly modify precision
 electroweak parameter $ S $ as discussed in ref.  \cite{Manohar:2006ga}.
  The Higgs boson production and decay $H \to \gamma \gamma$ were
 discussed \cite{Fajfer:2013wca,Cao:2013wqa}. For masses of color
 octet scalar $\sim 1 {\rm TeV}$ these modifications are negligible. 
On the other hand,  low energy flavor physics gets contribution from
 virtual colored scalar states. The contribution proportional to 
the quadratic coupling $\eta_U^2$ enters in $K^0 -\bar K^0$ mixing quantities
and should therefore be very small \cite{Manohar:2006ga}.
  The $\eta_D$ coupling appears in $B$ physics,  in particular 
a contribution proportional to a product of $\eta_U \, \eta_D$ gives
 rise to weak radiative $B$-meson decays.
In ref.  \cite{Manohar:2006ga,Trott:2010iz} it was  observed  that the phase of 
$\eta_U \eta_D$ can contribute to the 
CP violating asymmetry in $D$-decays.
 
  Following 
\cite{Isidori:2011qw} one can write an effective Hamiltonian
for non-leptonic charm decays
\begin{eqnarray} 
{\cal H}_{|\Delta C=1|} &=& \frac{G_F}{\sqrt 2}
 \sum _{i=1,2,5,6} \sum_{q} (C_i^q Q_i^q + C_i^{q'} Q_i^{q'} ) +  
 \frac{G_F}{\sqrt 2} \sum _{i=7,8} (C_i Q_i +C_i^{'} Q_i^{'})  + h. \, c.\, .
\label{e5}
 \end{eqnarray} 
 The full expression for the quark operators $Q_i$ are 
given in \cite{Isidori:2011qw}.
Among all possibilities it was found in \cite{Isidori:2011qw,Isidori:2012yx}
 that most likely, a  candidate to explain discrepancy between the experimental
 result and the SM 
prediction is the color-magnetic dipole operator 
$Q_8= \frac{m_c}{4 \pi^2}\bar u_L \sigma_{\mu \nu} T^A g_s G^{\mu \nu}_A c_R$. 
The same authors
found that NP is entering into the   Wilson coefficient $C_8$  such that
$\Delta a_{CP}^{exp} = - 1.8 \; {\rm Im}  \, C_8^{NP}(m_c)$. 
Using universal couplings $\eta_U$ and $\eta_D$, it was found
 \cite{Manohar:2006ga,Trott:2010iz} 
 that the bounds on
 these couplings
  $  {\rm Im} ( \eta_U^* \eta_D^*)$
 determined 
from $b \to s \gamma$ give a NEDM consistent with the current experimental
 bound, 
while we found that the corresponding parameters constrained by the
 direct CP violation in the charm sector given by Eq. (\ref{e1})
 are already ruled out  by the experimental result on NEDM.
In  ref. \cite{Manohar:2006ga} the most important contribution to the NEDM was 
found to arise from Weinberg's CP-violating three-gluon operator  operator.
 This contribution is generated by 
 the CEDM of the $b$-quark when
it is integrated out to obtain  the Weinberg operator. 
This means that the $b$ quark EDM
inducing Weinberg's
 operator gives  bounds on $  {\rm Im} ( \eta_U^* \eta_D^*)$. 
 These bounds are strong enough to exclude an explanation of the  CP violation
 in charm physics. 
 
 Therefore, the explanation of the  direct CP violation in $D$ decays
 by the presence of the color dipole operator, induced by the color 
octet scalar, is already excluded. 
 A viable possibility is to deviate from the generation universality 
for the color octet couplings to the quark fields, or to  modify the 
original MFV set-up of  \cite{Manohar:2006ga,Trott:2010iz}. 
Thus,  the authors of \cite{Altmannshofer:2012ur}  suggested small
 deviation from the MFV ansatz 
by allowing flavor changing $u \leftrightarrow c$ quark 
interactions with a neutral color octet scalar. 
Following ref. \cite{Altmannshofer:2012ur} 
 the interaction Lagrangian  is given by:
\begin{equation}
  {\cal L}_\mathrm{eff} \, = \, 
G(c \rightarrow u) \bar{u}_L \, T^A \, \Phi^A \, c_R \, 
+ \, X_d \bar{d}_L \,t^A  d_R \, \Phi^A \; + \; h.c. \; \, ,
\label{FCNC}
\end{equation}
where 
the couplings $G(c \rightarrow u)$ and $X_d$ are proportional
 to quark masses: 
\begin{equation}
G(c \rightarrow u) \equiv [X_u]_{12} = \zeta_u \, y_c \, X_{cu} \; ; \quad  
 X_{cu} \sim V_{cs} V_{us}^* \quad  ; \quad X_d = \zeta_d \, y_d \; \, ,
\label{couplings}
\end{equation}
where $\zeta_{u,d}$ are numbers (-to be determined by CP-violation 
in the charm sector) and $y_q =m_q/v$, where $v$ is the VEV of the Higgs,
and $m_q$ is the mass of quark $q$. 
In this framework the 
$D^0 -  \bar D^0$ mixing are not present at  tree level. 
In the case of $D^0 \to \pi^+ \pi^-$, the effect is negligible in
 comparison with 
 the $D^0 \to K^+ K^-$ amplitude  
 due to the smallness of the down
 quark mass compared to the bigger mass of the strange quark.
In the $D^0 \to K^+ K^-$ decay amplitude,  it was found that 
two operators  $\tilde O_{S1}^1 = ( \bar u P_R s) ( \bar s P_R c)$ and 
 $\tilde O_{S2}^1 = ( \bar u_\alpha P_R s_\beta) ( \bar s_\beta P_R c_\alpha )$ 
contribute to the effective Hamiltonian as described
 in \cite{Altmannshofer:2012ur}. 
Motivated by the model of  \cite{Altmannshofer:2012ur}, we use this tree 
level effective $(c \, u \,\Phi_8)$ coupling
 bounded by the 
 charm CP asymmetry.  Then we find that there is a new two-loop contribution,
 which we consider in the following subsection. We also find that these
 couplings induce higher dimensional operators present in NEDM. 
 It is important to notice that such  couplings cannot affect any other
 low energy observable, as already discussed in 
ref. \cite{Altmannshofer:2012ur}. 
The Barr-Zee mechanism induced by the color octets has already been
 considered in ref. \cite{Heo:2008sr}  for the EDM of the electron.

\subsection{Two-loop contributions to EDM} 

Two-loop contributions are induced by the presence of the $c \to u$ 
flavor changing
 color octet couplings described in the modified MFV framework above. 
In our calculation we use  the effective fermion (quark) propagator in a soft 
gauge field \cite{Reinders:1984sr}:
\begin{equation}
S_1(k,F) \; = \; (- \frac{e_q}{4}) \;
 \frac{ \left\{(\gamma \cdot k + m_q) \, , \, \sigma \cdot F \right\} }
{(k^2 - m_q^2)^2}\,, 
\label{propagator}
\end{equation}
where $F$ is the electromagnetic field,  $\{A,B \}$ denotes the
 anti-commutator,
 $k$ is the four momentum and $m_q$ the mass of the quark $q$.

\begin{figure}[t]
\begin{center}
    \epsfig{file=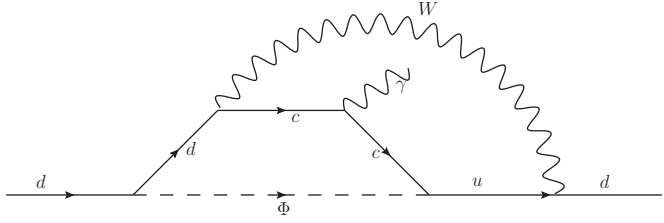,width=9cm}
\caption{Two loop diagram for an EDM of a $d$-quark.}
\label{fig:2loop}
\end{center}
\end{figure}

\begin{figure}[t]
\begin{center}
    \epsfig{file=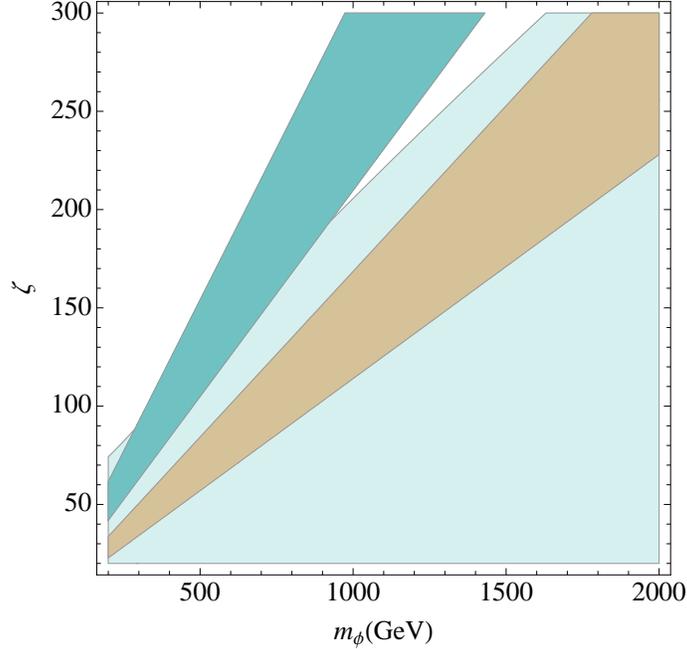,width=9cm}
\caption{Regions in the $\zeta - M_\Phi$ plane compatible with the data on 
$\Delta a_{CP}$ (dark green, $C_H=1$ and pale  brown for $C_H \simeq 3$) and  on the current experimental lower bound on $NEDM$ (pale green). }
\label{fig:NEDM-ACP}
\end{center}
\end{figure}

We have found that the two loop contribution in Fig. \ref{fig:2loop}
induces a dimension 5  electric dipole operator for the $d$-quark
 which might be  written as an effective Lagrangian in the following way:
\begin{eqnarray}
{\cal L}_1(d \rightarrow d \gamma)_\Phi \; =
 \; K \, \left(\bar{d}_L \, \sigma \cdot F \, d_R \right). 
 \label{e-Two-loop}
\end{eqnarray}
The quantity  $K$ is given by 
\begin{eqnarray}
K \, = \, C_3  \, \lbrack g_W^2 V_{ud} \, V_{cd}^* \,
 G(c \rightarrow u) X_d \rbrack  \, 2 m_c \, e_c \; I_{2-loop} \; ,
\end{eqnarray}
where $C_3 \equiv < T^A \, T^A > = 4/3$ is a color factor and 
the leading logarithmic approximation of the two loop integral is:
\begin{eqnarray}
I_{2-loop} \, \simeq \, \left( \frac{1}{16 \pi^2}\right)^2 \, \frac{1}{M_\Phi^2}
\; 
 \left(\left[ln\frac{M_\Phi^2}{m_c^2}\right]^2
 \, - \, \left[ln\frac{M_W^2}{m_c^2}\right]^2 \, \right) \; \, .
\label{eq:2loop-int}
\end{eqnarray}
There is also a contribution from the crossed diagram with the result:
\begin{eqnarray}
{\cal L}_2(d \rightarrow d \gamma)_\Phi \; =
 \; K^*  \, \left(\bar{d}_R \, \sigma \cdot F \,  d_L \right), 
  \, 
\label{eq:Two-loopC}
\end{eqnarray}
such that there will be an EDM
 of the $d$-quark equal to $(d_n)_{2-loop}^\Phi = 2 Im(K)$. 
Note that these diagrams appear due to the presence
 of appropriate chiralities in the interacting Lagrangian (\ref{FCNC}). 
The EDM
 of the $u$-quark
is suppressed by extra factors of small masses.
We have checked that
 the opposite chirality of the the scalars  lead to the two-loop 
 helicity suppressed amplitudes. 
  Following  the  work of  \cite{Altmannshofer:2012ur},  we have found that
 one can write 
the  asymmetry in Eq. (\ref{e1}), assuming maximal phase $\Phi_f$, as
\begin{eqnarray}
\Delta a_{CP} = \frac{2}{9}\frac{\zeta^2}{M_{\Phi}^2} \, m_K^2 \,
  C_{RGE} \,C_H \,
\label{eq. Alt-ACP}
\end{eqnarray}
 where $C_{RGE}$ denotes the factor which
 includes running of the Wilson
 coefficient  ($C_{RGE} =0.85$, for the running from the scale
 $M_\Phi \sim 1 \,$  TeV  down to the scale equal $m_c$), while 
 $C_H$ stands for the  possible enhancement of the
 hadronic matrix elements,  assumed to be as large as $C_H \simeq 3$
 in comparison with 
 the naive factorization estimate leading to $C_H =1$ as 
explained in   \cite{Altmannshofer:2012ur}. 
 We denote $\zeta^2 = \zeta_u \, \zeta_d$, which appears in both
 expressions for NEDM and $\Delta a_{CP}$. 
Assuming that our result for the $d$ quark electric dipole moment gives
 arise to NEDM  as given in (\ref{valence}), we  bound our  new
 contribution 
 to NEDM by the current experimental result given in (\ref{e3}). 
Both constraints are presented on Fig.~\ref{fig:NEDM-ACP}.  

In the case that the mass of color octet is bounded in the TeV 
regime  \cite{ATLAS:2012ds,CMS:2012yf}, the parameter $\zeta^2$ should
 be scaled accordingly. We have checked that even for 
a mass of $M_\Phi \geq 1.86$ TeV, perturbativity
is still valid for  
the couplings in (\ref{FCNC}).  
That means that one can safely use the effective 
Lagrangian (\ref{FCNC}) for this purpose. 
We point out that  the loop diagram in 
Fig.~\ref{fig:2loop} is finite due to the chiral structure of
(\ref{FCNC}), and that we have only given the 
 relevant leading-log result. 
Of course the leading logarithmic result in (\ref{eq:Two-loopC}) 
will be modified when taking into account higher orders by means of
  renormalization group techniques, but at the present stage we use
 this result.
 For fixed asymmetry - i.e. $\frac{\zeta^2}{M_\Phi^2}$ fixed -
we obtain the relation
\begin{eqnarray}
(d_n/e)_{2-loop}^\Phi \; \, \simeq \, 
\left(\frac{\lambda^2 \, m_d}{8 \pi^4}\right) \, 
\frac{M_W^2 m_c^2}{v^4 \, m_K^2} \, \frac{\Delta a_{CP}}{ C_{RGE} \,C_H }
\,
 \left(\left[ln\frac{M_\Phi^2}{m_c^2}\right]^2
 \, - \, \left[ln\frac{M_W^2}{m_c^2}\right]^2 \, \right) \; \, .
\label{eq:dnaCP}
\end{eqnarray}
Numerically,
we obtain the range 
\begin{equation}
(d_n/e)_{2-loop}^\Phi \; \simeq  \; (1.0 - 2.3) \times 10^{-26} \, 
\mbox{cm} \; \,  ,
\label{D-2l}
\end{equation}
for $M_\Phi$ in the range 400 GeV to 2 TeV,  assuming $C_H\simeq 3$.
 It is interesting that  for $C_H =1$, we would get
$d_n$ three times bigger and violate the experimental 
bound in Eq.(\ref{e3}) ! 

Let us comment that this result should also be valid for other flavor 
changing $(c \, u  \, \Phi)$ colored 
scalar (triplet, sextet)  couplings, 
with the appropriate color factor replacement.

 \subsection{ Higher-dimensional operators} 
 
 We now  consider the di-quark mechanism and calculate one loop diagrams
for $W^- \, u  \rightarrow d \gamma$ and $W^+ \, d  \rightarrow u \gamma$
and afterwards attach  left-handed
 $u \rightarrow d$ and $d \rightarrow u$ currents.
\begin{figure}[t]
\begin{center}
    \epsfig{file=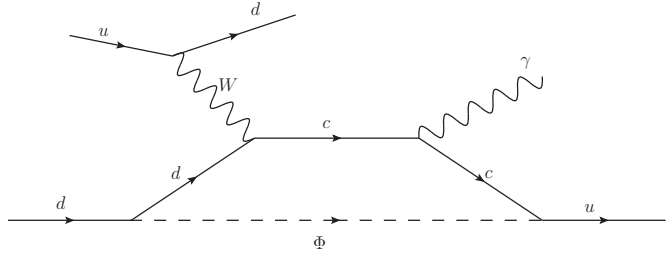,width=9cm}
\caption{One loop diagram for $W^+ \, d  \rightarrow u \gamma$.}
\label{fig:1loop}
\end{center}
\end{figure}
The contribution from the one loop diagram $W^+ \, d  \rightarrow u \gamma$
(from Fig.~\ref{fig:1loop} ), leading to di-quark mechanism 
when $W$ is connected  to a left-handed current, is
\begin{eqnarray}
{\cal L}(d u \rightarrow u \, d \gamma) \; = \; C_3\,
 \lbrack \frac{e_c \, g_W^2}{4 M_W^2} V_{ud} \, V_{cd}^* \, 
G(c \rightarrow u) X_d\rbrack
 \, 2 m_c \, \; I_{1-loop} \, Q(u d \rightarrow d \, u \, \gamma)_\Phi \; \,  ,
\label{eq:One-loop}
\end{eqnarray}
where $C_3$ is the color factor defined above, 
$g_W$ is the $W$ coupling,
\begin{eqnarray}
 Q(u d \rightarrow d \, u \, \gamma)_\Phi \; = \; 
\left(\bar{u} \, \sigma \cdot F \, \gamma_\mu L \, \gamma^\nu \, d \right)
i D^\nu \left(\bar{d} \gamma^\mu L \, u \right)
\label{eq:One-loop-op}
\end{eqnarray}
is a generated dimension 9 operator, and 
\begin{eqnarray}
I_{1-loop} \, \simeq \,  \frac{1}{16 \pi^2} \, 
\frac{1}{m_c^2 M_\Phi^2} \; \, .
\label{eq:1loop-int}
\end{eqnarray}
The four quark operator part of $Q(u d \rightarrow d \, u \, \gamma)_\Phi$
 is of the same type as obtained in eq. (\ref{edq}) 
and can be estimated by using, say an
 $N^*$ resonance between the two currents. The covariant derivative,
 corresponding to the $W$-momentum,
 will contribute with a momentum  of order
 the constituent quark mass in nucleons, i.e.
 $\hat{m} \equiv  m_{constit} \simeq$ 350 MeV.
There are also 3 more relevant diagrams, leading to a NEDM
\begin{equation}
d_n  \; \sim \;   \, C_3 \,
[\frac{(e_c -e_d) \, g_W^2}{4 M_W^2} V_{ud} \, V_{cd}^* \, G(c \rightarrow u) X_d]
 \, 2 m_c \, \; I_{1-loop} \,  F_{Hadr} \; ,
\label{dn1loop}
\end{equation}
 where 
$ F_{Hadr}$ is a pure  hadronic factor 
\begin{equation}
 F_{Hadr} \, = \, \hat{F} \, \hat{m} \, M_* \, I_{sp} \; ,
\end{equation}
where
$I_{sp}$ is the phase space integral for the intermediate
 $N^*$ (with mass $M_*\simeq 1.440$ GeV) 
 found to be
  $I_{sp}  \simeq 0.9 \cdot 10^{-2}$ GeV$^2$. 
 The quantity $\hat{F}$ is a product of transition 
form factors
 (from $N^*$ to $N$), which takes care of the damping within these,  
and is expected to be between 0.3 and 0.6.
We  find 
\begin{equation}
d_n/e \; \simeq  \; 3 \times 10^{-31} \, \mbox{cm} \; , 
\label{NEDM-4q}
\end{equation}
 which is a bit higher than the corresponding SM value 
 based on the operator in (\ref{edq}).
 As noticed by  \cite{Mannel:2012qk}  one has to be careful
 in neglecting higher dimension operators. 

\section{Conclusions}
We have investigated contribution to the NEDM  induced by the presence of the 
non-MFV flavor changing ($c\,u\,\Phi_8$) coupling. 
 The relevant couplings can be  constrained by the world average CP asymmetry 
found in $D\to K^+ K^-/\pi^+ \pi^-$ decays or by current experimental bound 
on NEDM.  
It is remarkable that the CP violating asymmetry in charm  decays and
our  new 
 NEDM  two-loop contribution  allow sizable  parts of the parameter space
 in the $\zeta- M_\phi$ plane. 
Still the CP violating asymmetry is more constraining than the bound on NEDM. 
We comment on the higher dimensional operator contribution to NEDM and 
found that this contribution might be $10$ to $10^2$ larger than the
 NEDM values within the SM.
Our study is applicable for flavor non-diagonal  $(c\,u\,)$ couplings
 to a  scalar,
 by replacing the color factor in the amplitude. 
 
The further LHC searches might set new bounds on the masses of colored 
scalars.  On the other hand,   new measurements of the    CP asymmetry 
in charm physics  would
 shed more light on the 
possible new source of CP violation in flavor physics.
At the same time improvements of the experimental value for the NEDM might help
 in clarifying the role of new physics  induced CP violating phases. 

{\acknowledgments}

JOE is supported in part by the Norwegian
  research council. 
The  work of SF is partially supported by Slovenian research agency ARRS.

\end{document}